\documentclass[prl,twocolumn,showpacs,nobibnotes]{revtex4}

\usepackage{bm}
\usepackage[dvips,final]{graphics}

\begin{document}

\title{Staggered Flux Phase in a Model of Strongly Correlated Electrons}

\author{J. B. Marston} 
\author{J. O. Fj{\ae}restad}
\affiliation{Department of Physics, Brown University, Providence, RI
02912-1843}

\author{A. Sudb{\o}} 
\affiliation{Department of Physics, Norwegian University of Science
and Technology, N-7491 Trondheim, Norway}

\date{February 11, 2002}  

\begin{abstract}
We present numerical evidence for the existence of a staggered flux (SF) 
phase in the half-filled two-leg t-U-V-J ladder, with true long-range order 
in the counter-circulating currents.  The density-matrix renormalization-group 
(DMRG) / finite-size scaling approach, generalized to describe complex-valued 
Hamiltonians and wavefunctions, is employed.  The SF phase exhibits robust 
currents at intermediate values of the interaction strength.
\end{abstract}


\maketitle

\begin{figure}
\resizebox{8cm}{!}{\includegraphics{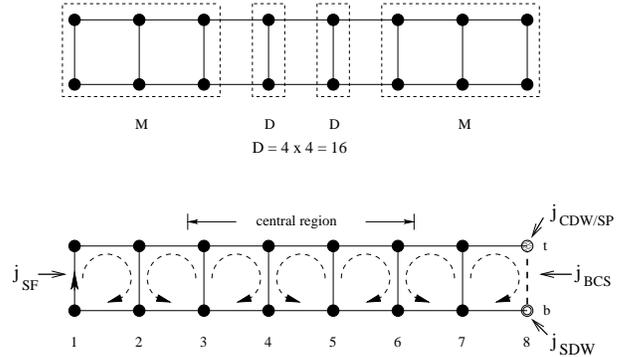}}
\caption{Upper diagram: A two-leg ladder of length $L = 8$.
The left and right blocks each retain Hilbert spaces of dimension $M$.  
To implement the DMRG / finite-size algorithm, the two sites belonging
to each rung are paired together into a single site with a Hilbert space
of complex dimension $D = 16$.  Lower diagram: Source current $j_{SF}$ applied to the
left-most rung induces currents in the interior of the ladder
(dashed circles). A source of Cooper pairs ($j_{BCS}$), a local chemical
potential ($j_{CDW/SP}$), and a local magnetic field ($j_{SDW}$) are added 
at the ladder ends to induce BCS, SP, CDW, and SDW order in the central
region of the ladder, which is then monitored as the ladder length increases.} 
\label{fig:ladder}
\end{figure}

The zero-temperature phase diagram of the two-dimensional Hubbard
model and its various extensions remains poorly understood.  An
intriguing possibility that has been the focus of considerable
attention is known variously as the ``orbital
antiferromagnet''\cite{halperin,OAF1,OAF2}, the ``staggered flux''
(SF) phase\cite{AM,MA,ted,ivanov,leung,didier}, or the ``d-density
wave''\cite{chetan,nayak}.  The state breaks both time-reversal 
and lattice translation symmetries; another phase, 
the ``circulating current phase''\cite{varma}, is similar but
preserves translational symmetry.  Such ``hidden'' forms
of order could arise in the pseudogap phase of the cuprate
superconductors, implying the existence of a quantum critical point
and possibly also non-Fermi liquid behavior\cite{nayak,varma}. The SF
phase competes against other better-known phases such as
charge-density waves (CDW), spin-density waves (SDW),
superconductivity (BCS), stripes, and Fermi liquids.  It is important
to ascertain whether or not such order can really occur in sensible
microscopic models of correlated electrons.
In this Letter we study numerically the
simplest possible system that could support such a phase, namely an
extended version of the Hubbard model on a two-leg ladder (see Fig.
\ref{fig:ladder}).  Our approach extends to arbitrary interaction strength
earlier analytical work by two of us that showed that the SF phase arises
at weak-coupling\cite{john}.  

Ladder systems are of interest both because they
are easier to understand theoretically than full two-dimensional lattices,
and also because they are realized in nature\cite{dagotto}. Although
ladders are one-dimensional, true long-range order (LRO) in the
orbital currents is possible at zero temperature because the currents
break discrete symmetries. This is in contrast to spin-density wave or
BCS superconducting order which break continuous symmetries and which
therefore can at most exhibit quasi-LRO with power-law decay. 

Away from half-filling, numerical studies of the t-J
model\cite{SWA,Asle} and various extensions of it\cite{troyer} on the
two-leg ladder found no evidence for a SF phase. However, for weak
interactions, tendencies towards SF ordering (not true LRO, but
rather power-law decay of current-current correlation functions) 
have been found analytically
both for the spinless\cite{ners-luther-kus} and spinful\cite{schulz96}
cases (see also Ref.~\onlinecite{orignac}).

Here we will consider the half-filled two-leg ladder. We study an
extended Hubbard model, the t-U-V-J model, defined on the two-leg
ladder as follows:
\begin{widetext}
\begin{eqnarray}
H &=& \sum_{i=1}^{L-1} \sum_{\lambda=t,b} \bigg{\{}
-t_\parallel (c^{\dagger \sigma}_{i+1,\lambda} c_{i,\lambda,\sigma} + H.c.)
+ V_\parallel (n_{i+1,\lambda} - 1) (n_{i,\lambda} - 1)
+ J_\parallel \vec{S}_{i+1,\lambda} \cdot \vec{S}_{i,\lambda}
\bigg{\}}
\nonumber \\
&+& \sum_{i=1}^L \bigg{\{}
-t_\perp (c^{\dagger \sigma}_{i,t} c_{i,b,\sigma} + H.c.)
+ V_\perp (n_{i,t} - 1) (n_{i,b} - 1) 
+ J_\perp \vec{S}_{i,t} \cdot \vec{S}_{i,b}
\bigg{\}} + \frac{U}{2}~ \sum_{i=1,\lambda}^L (n_{i,\lambda} - 1)^2\ . 
\label{Hamiltonian}
\end{eqnarray}
\end{widetext}
Here $c^{\dagger \sigma}_{i \lambda}$ creates an electron of spin
$\sigma = \uparrow, \downarrow$ on site $i$ of leg $\lambda = t, b$ of
the ladder.  Operators $n_{i,\lambda} \equiv c^{\dagger
\sigma}_{i,\lambda} c_{i,\lambda,\sigma}$ and $\vec{S}_{i,\lambda}
\equiv \frac{1}{2} c^{\dagger \alpha}_{i,\lambda}
\vec{\sigma}_\alpha^\beta c_{i,\lambda,\beta}$ are respectively the
number and spin at site $(i,\lambda)$.  There is an implicit sum over
repeated raised and lowered spin indices. Particle-hole symmetry
implies that the chemical potential $\mu = 0$ when the system is
half-filled, with $\langle n_{i,\lambda} \rangle = 1$ on each site.

For $V_\parallel = J_\parallel = 0$ and
$J_\perp = 4 (U + V_\perp)$ the model has SO(5)
symmetry\cite{SZH}.  The phase diagram of the SO(5)-symmetric model 
with $t_\perp = t_\parallel$ was investigated in Ref.~\onlinecite{LBF98}
in the weak-coupling limit using a perturbative renormalization-group (RG)
analysis combined with bosonization. One of the phases of the SO(5) model, 
located in the $U-V_{\perp}$ plane between the lines $V_{\perp} = -2 U$ and 
$V_{\perp} \approx -5.7 U$ with $V_{\perp}>0$ and $U<0$, was identified as
having spin-Peierls (SP) order. However, a recent bosonization study
by two of us\cite{john} revealed that this phase in actuality exhibits
no dimerization, but instead is the SF phase with true LRO in the
orbital currents. For a correct understanding of the problem it was
essential to treat carefully the ``Klein factors'' that must be
introduced to maintain anti-commuting statistics of the (bosonized)
fermionic degrees of freedom.

Previous RG studies have shown that in weak coupling a rather generic
ladder model flows to a manifold with SO(5) symmetry for a range of
values of the model parameters\cite{LBF98,arrigoni}. In agreement with
these results, we have shown that there are RG flows towards the SF phase
also when not too large non-SO(5)-invariant terms (such as $J_{\parallel}$ and
$V_{\parallel}$) are added to the SO(5)-model\cite{in_preparation}. 

Weak-coupling RG is unreliable at intermediate interaction strengths.  
Instead we use the ``infinite-size'' version of the density-matrix 
renormalization-group
(DMRG) algorithm\cite{white} to search for the ordered currents.  The
half-filled ladder is expected to be fully gapped over a large portion 
of the phase diagram; consequently (as shown
below) the infinite-size algorithm is sufficiently accurate to obtain
well-converged results.  Each site of the ladder has a Hilbert space
of dimension $4$, as the site can either be unoccupied, have a single
electron of either up or down spin, or be doubly occupied. To employ
the DMRG algorithm, we group pairs of sites connected by a rung into a
single site of Hilbert space dimension $D = 4 \times 4 = 16$ 
(see Fig. \ref{fig:ladder}).  Errors
in the calculation of observables introduced by the DMRG truncation of
the Hilbert space can be systematically reduced by increasing the size
$M$ of the Hilbert space retained in each of the two outer blocks up
to limits imposed by computer memory and speed.  The  
largest Hilbert space we use has $M = 150$ with corresponding complex
dimension $M \times D \times D \times M = 5,760,000$.  

Questions of the spontaneous formation of order are addressed by the
combined DMRG / finite-size scaling approach described in some detail
in Ref.~\onlinecite{tsai}.  Quantum-critical points have been studied with
the method, and critical exponents have been obtained at percent-level
accuracy\cite{senthil,tsai}.  In the present case we induce
symmetry breaking by applying a source current $j_{SF}$ to the left end
rung of the ladder.  Terms that induce Cooper-pair formation by
breaking U(1) particle-number symmetry ($j_{BCS}$), induce CDW and
SP order by breaking lattice reflection symmetries ($j_{CDW/SP}$), 
and induce a SDW through the application of a local magnetic field ($j_{SDW}$), 
are also added to the right end of the ladder (see Fig. \ref{fig:ladder}):
\begin{widetext}
\begin{eqnarray}
H \rightarrow H &+& 
j_{SF} * i t_\perp (c^{\dagger \sigma}_{1,t} c_{1,b,\sigma} - H.c.)
+ j_{BCS} * (c^{\dagger \uparrow}_{L,t} c^{\dagger \downarrow}_{L,b} 
- c^{\dagger \downarrow}_{L,t} c^{\dagger \uparrow}_{L,b} + H.c.)
\nonumber \\
&+& j_{CDW/SP} * n_{L,t} + j_{SDW} * S^z_{L,b}\ .
\end{eqnarray}
\end{widetext}
In addition to these explicit source terms, 
we note that the open boundary conditions on the ladder also
act as O(1) source terms for a columnar dimer pattern\cite{tsai}. 
As the Hamiltonian is complex-valued when $j_{SF} \neq 0$, we
generalize the DMRG code to handle complex-valued wavefunctions and
reduced density matrices.  This generalization comes at the cost of
doubling the required memory, and slowing down both the sparse and the
dense matrix diagonalization, but the cost is offset by the advantage
that now we can access fully ordered ground states, well beyond linear
response theory.

We calculate the expectation value of the current operators $2
t_\perp~ {\rm Im}\{c^{\dagger \alpha}_{i,t} c_{i,b,\alpha}\}$ and $2
t_\parallel~ {\rm Im}\{c^{\dagger \alpha}_{i+1,\lambda}
c_{i,\lambda,\alpha}\}$ respectively on the central rungs and links of
the ladder, checking that current is conserved (Kirchhoff's junction
rule) at the vertices in the central region; deviations are typically of
order $10^{-11}$ in units where $t_\perp = t_\parallel = 1$.  We also monitor the 
Cooper pair amplitude, $\langle c^{\dagger \uparrow}_{i,t} c^{\dagger \downarrow}_{i,b} 
- c^{\dagger \downarrow}_{i,t} c^{\dagger \uparrow}_{i,b} \rangle$; 
the average magnitude of deviations in the electron hopping amplitudes 
$\langle c^{\dagger \sigma}_{i, \lambda} c_{j, \lambda^\prime, \sigma} 
+ H.c. \rangle$ from their mean value; deviations in the electron occupancy from one, 
$\langle n_{i,\lambda} - 1\rangle$; 
and the local spin density, $\langle S^z_{i,\lambda} \rangle$.  

\begin{figure}
\resizebox{8cm}{!}{\includegraphics{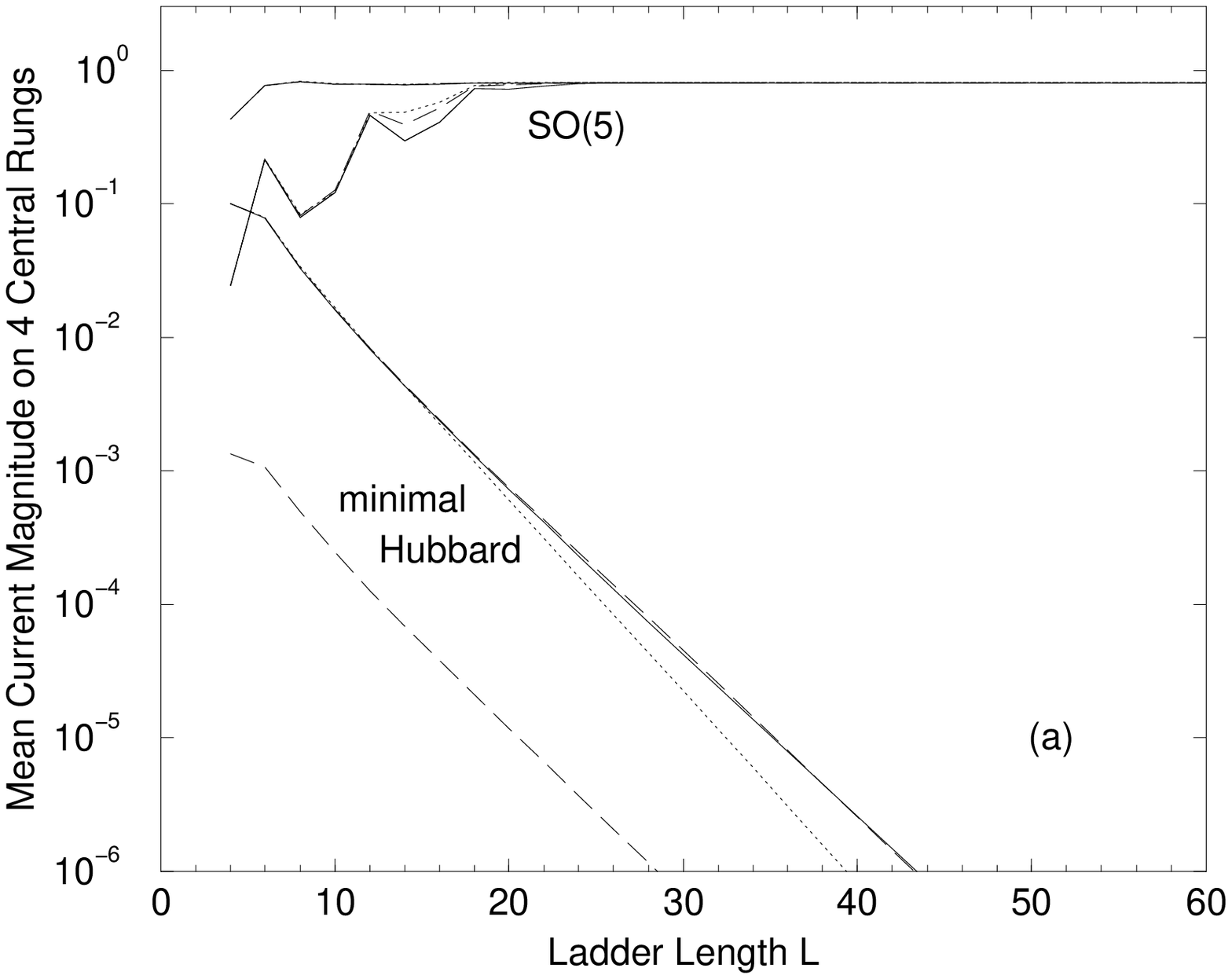}}
\resizebox{8cm}{!}{\includegraphics{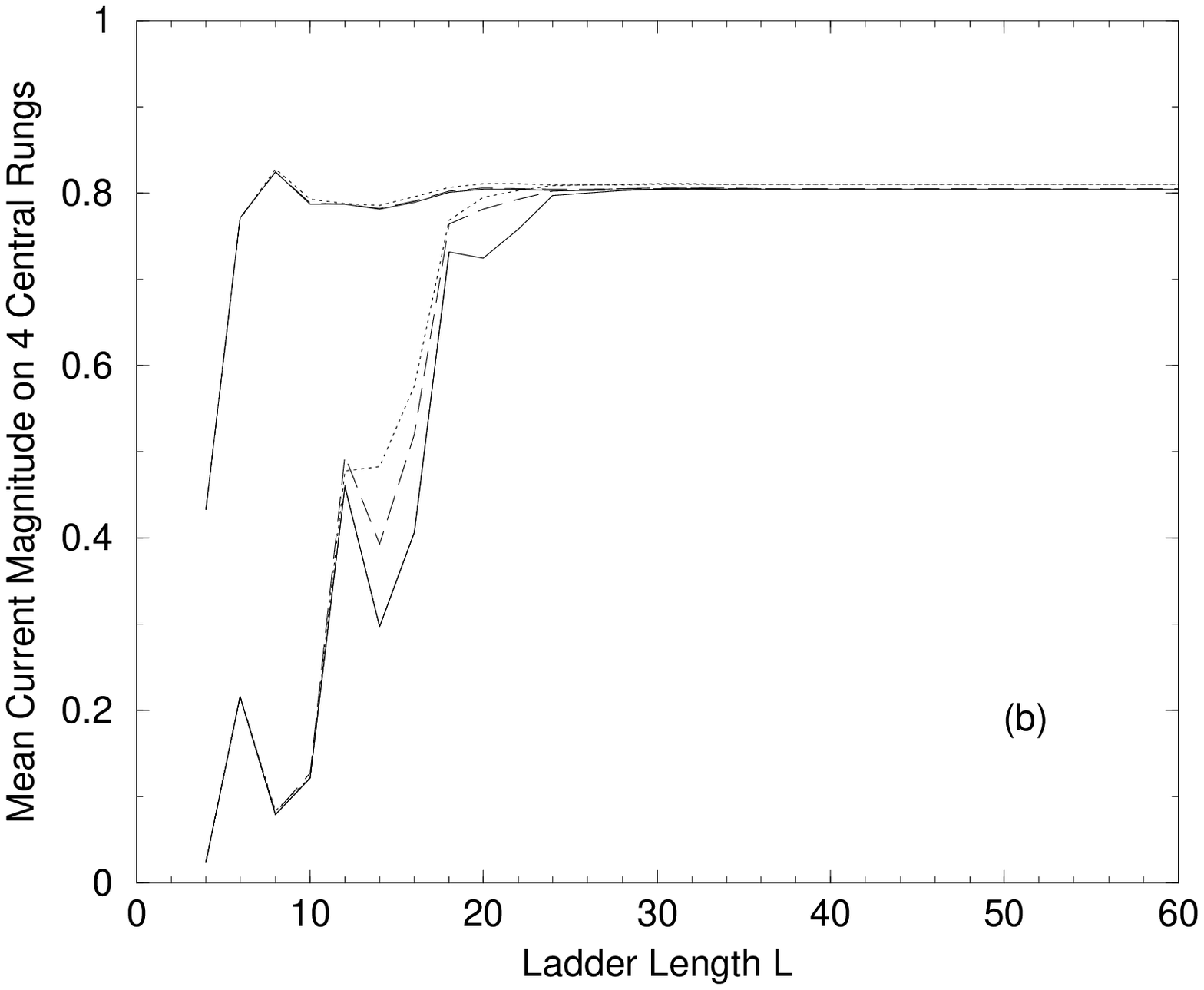}}
\caption{Semilog (a) and linear (b) plots of the magnitude of the induced current 
averaged over the four central rungs of the ladder. 
Dotted lines are for calculations with $M = 50$,
dashed lines are for $M = 100$, and solid lines are for $M = 150$.
Results are shown for the case of the SO(5) ladder with 
$t_\perp = t_\parallel = 1$, $U = -0.4$, $V_\perp = 0.9$, $J_\perp = 2$, and $V_\parallel =
J_\parallel = 0$.  
Also shown on the semilog plot is the case of the minimal Hubbard model with 
$U = 4$ and $j_{SF} = 0.01$ (lower line that drops exponentially) and 
$j_{SF} = j_{BCS} = j_{CDW/SP} = j_{SDW} = 1$ (upper lines that drop exponentially). 
In both plots, the lower set of SO(5) curves is for 
a small source current $j_{SF} = 0.01$ applied to the ladder's left edge. 
The upper set is for large source currents $j_{SF} = j_{BCS} = j_{CDW/SP} = j_{SDW} = 1$. 
Note the convergence to the same asymptotic value of current 
as the thermodynamic limit of long ladder length is approached.  The asymptotic current
for $M=100$ differs by 0.6\% from that at $M=50$; the value for $M=150$
differs by only $0.1\%$ from that at $M=100$ demonstrating good convergence.}  
\label{fig:currents}
\end{figure}

First consider the pure minimal Hubbard t-U model with $t_\perp =
t_\parallel = 1$, $U = 4$, and $V_\perp = V_\parallel = J_\perp =
J_\parallel = 0$.  As shown in the semilog plot of Fig. \ref{fig:currents}(a), 
applied source currents of $j_{SF} = 1$ and $0.01$ induce currents on the four
central rungs of the ladder that decrease exponentially  
as the ladder is enlarged via the DMRG algorithm, in agreement with the
weak-coupling RG calculation\cite{LBF98}.  Likewise we find no
instabilities towards BCS, SP, CDW or SDW order.  Instead the phase is a
fully-gapped insulator qualitatively the same as that found in the 
Heisenberg antiferromagnet. 

\begin{figure}
\resizebox{8cm}{!}{\includegraphics{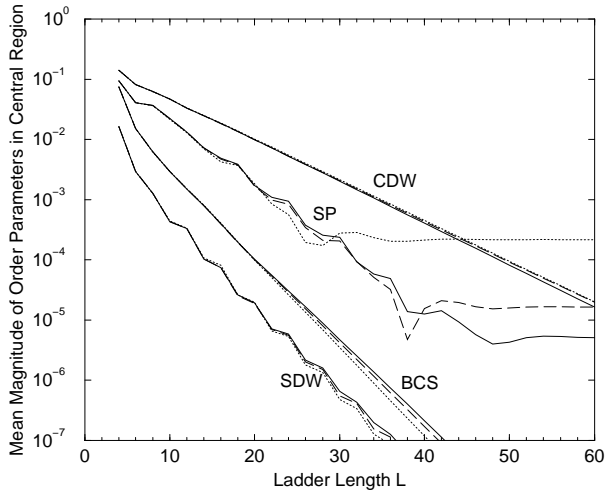}}
\caption{Semilog plot of various order parameters in the SF phase (see text).
Edge source terms $j_{BCS} = j_{CDW/SP} = j_{SDW} = j_{SF} = 1$ 
induce modulations in corresponding observables at the center of the ladder.  
Dotted, dashed, and solid lines correspond to $M = 50$, $100$ and $150$ respectively.
Plotted in the figure, from highest to lowest in magnitude, are the order parameters 
for CDW, SP, BCS, and SDW tendencies.  See text for the definition of these
order parameters.  
Each of the expectation values decays exponentially with increasing ladder length.  
(The SP order parameter eventually levels off, but this residual order vanishes
as the block size $M$ is increased, showing that it is an artifact of the Hilbert
space truncation.) 
Similar results are obtained upon applying either large or small source terms, one
at a time. 
Thus there is no tendency towards BCS, SP, CDW or SDW order in the SF phase.}
\label{fig:otherorder}
\end{figure}

Next consider the effect of turning on interactions $J_\perp$ and
$V_\perp$ along the rungs of the ladder.  According to the phase diagram in 
Ref. \onlinecite{LBF98}, for the case of exact SO(5) symmetry with $J_\perp = 4 (U + V_\perp)$
the SF phase should occur\cite{john} in the weak-coupling limit 
for $U < 0$ at $V_\perp / U = -9/4$. 
Fig. \ref{fig:currents} shows that the SF phase does indeed arise at 
these ratios for intermediate interaction strength:  For
$U = -0.4$, $V_\perp = 0.9$, and $J_\perp = 2$, a small source current induces
orbital currents in the central region of the ladder that grow with increasing
ladder length.   The currents saturate to the same non-zero asymptotic 
value in the limit of large ladder length, regardless of the size of the 
source terms.  There is good convergence with increasing block Hilbert
space dimension $M$; the asymptotic value of the rung current 
for the most accurate truncation of $M=150$ is 0.8047 and differs by only 0.1\%
from the value obtained with $M=100$.  The currents alternate in sign from
rung to rung as expected in the SF phase.  There are no other
instabilities: As shown in Fig. \ref{fig:otherorder} 
the BCS, SP, CDW, and SDW order parameters all vanish exponentially 
as the ladder is enlarged via the DMRG algorithm, consistent with the
weak-coupling prediction.
The excitation spectrum remains fully gapped; consequently the SF phase
exists over a range of parameter space well beyond the region of 
exact SO(5) symmetry.  For example, robust currents arise 
for $U = -0.5$, $V_\perp = 0.75$, and $J_\perp = 2$,
and also when interactions along the links are turned on, $U = -0.5$, 
$V_\perp = V_\parallel = 0.75$, $J_\perp = 2$, and $J_\parallel = 0.5$. 
 
In the large-$J_\perp$ limit, with ratios $U / J_\perp = -1/4$ 
and $V_\perp / J_\perp = 3/8$ kept fixed, 
the ground state of the ladder consists of a direct product
of spin-singlet dimers on each of the rungs, with no broken
symmetries\cite{SZH}.  We have verified that the orbital currents 
cease when the interaction strength is increased further to
$J_\perp > 2.5$.  We are carrying out a more extensive investigation
of the phase diagram of model Eq. \ref{Hamiltonian} by using the
weak-coupling RG equations in combination with the DMRG / finite-size scaling
method\cite{in_preparation}.

It would be interesting to study the effects of doping away from half-filling.
Numerical study of the doped SF phase would require the use
of the more accurate ``finite-size'' DMRG algorithm, 
as either gapless excitations, or rich 
spatial structures such as stripes, are expected to occur\cite{WAS}.  
One question to be answered is whether or not LRO in the currents can 
still arise at commensurate hole concentrations. 

In summary we have shown that a phase of strongly correlated electrons
exists in which currents form spontaneously in the half-filled two-leg ladder.
At half-filling the SF phase is fully gapped, and thus is an insulator of
the Mott-Hubbard type.  Our results, suitably generalized to two or three 
spatial dimensions, may have application to several stoichiometric compounds that were
recently proposed to be in the SF phase\cite{stoichiometric}.
It is also intriguing that the SF phase lies in between a checkerboard CDW phase, 
and the D-Mott phase\cite{LBF98,john,in_preparation}, as
charge segration into stripes and d-wave superconductivity are two phases
that occur in the high-T$_c$ cuprates.

{\it Note added:} U. Schollw{\"o}ck has recently used the ``finite-size'' version
of the DMRG algorithm, generalized to complex-valued wavefunctions, to study the
SO(5) ladder with the same parameters as Fig. \ref{fig:currents}. 
The value of the saturated rung current that he finds agrees quantitatively with our result. 

We thank I. Affleck, L. Balents, S. Chakravarty, S. Kivelson,
D.-H. Lee, P. Lee, D. Scalapino, U. Schollw{\"o}ck, M. Troyer, 
S.-W. Tsai, C. Varma, and S. White for helpful discussions.  This
work was supported in part by the NSF under grant Nos. DMR-9712391,
CDA-9724347, and PHY99-07949.  
J.O.F. was supported by the Norwegian Research Council,
Grant No 142915/432, and A.S. by Grant No. 115541/410
Some computational work was carried out at Brown University's
Technology Center for Advanced Scientific Computing and Visualization.

\end{document}